\input ./style/arxiv-general.cfg
\documentclass[aoas,MSNbibl,nameyear,dvips]{arximspdf}
\makeatletter
   \@ifpackageloaded{graphicx}{}{\usepackage{graphicx}}
\makeatother
\usepackage{mathbh}
\usepackage{euromarv}

\doi{10.1214/15-AOAS823}
\volume{9}
\issue{2}
\pubyear{2015}
\firstpage{1024}
\lastpage{1052}
\docsubty{FLA}


\begin{document}
\begin{frontmatter}

\title{Bayesian structured additive distributional regression with an
application to regional income inequality in Germany}
\runtitle{Bayesian structured additive distributional regression}

\begin{aug}
\author[A]{\fnms{Nadja}~\snm{Klein}\corref{}\ead[label=e11]{tkneib@uni-goettingen.de}\thanksref{M1,T1}},
\author[A]{\fnms{Thomas}~\snm{Kneib}\thanksref{M1,T1}\ead[label=e1]{nklein@uni-goettingen.de}},
\author[B]{\fnms{Stefan}~\snm{Lang}\thanksref{M2,T2}}
\and
\author[A]{\fnms{Alexander}~\snm{Sohn}\thanksref{M1}}
\runauthor{Klein, Kneib, Lang and Sohn}
\affiliation{Georg-August-University G\"{o}ttingen\thanksmark{M1} and
University of Innsbruck\thanksmark{M2}}
\address[A]{N. Klein\\
T. Kneib\\
A. Sohn\\
Georg-August-University G\"{o}ttingen\\
Chair of Statistics\\
Platz der G\"{o}ttinger Sieben 5\\
37073 G\"{o}ttingen\\
Germany\\
\printead{e1}}
\address[B]{S. Lang\\
University of Innsbruck\\
Department of Statistics\\
Universit\"{a}tsstrasse 15\\
6020 Innsbruck\\
Austria}
\end{aug}
\thankstext{T1}{Support by the German Research Foundation via the
research training group 1644 and the projects KN 922/4-1/2 is
gratefully acknowledged.}
\thankstext{T2}{Supported by the Oesterreichische Nationalbank,
Anniversary Fund, project number: 15309.}

%
\received{\smonth{9} \syear{2013}}
%
\revised{\smonth{3} \syear{2015}}

%
\begin{abstract}
We propose a generic Bayesian framework for inference in distributional
regression models in which each parameter of a potentially complex
response distribution and not only the mean
is related to a structured additive predictor. The latter is composed
additively of a variety of different functional effect types such as
nonlinear effects, spatial effects, random coefficients, interaction
surfaces or other (possibly nonstandard) basis function
representations. To enforce specific properties of the functional
effects such as \mbox{smoothness}, informative multivariate Gaussian priors
are assigned to the basis function coefficients. Inference can then be
based on computationally efficient Markov chain Monte Carlo simulation
techniques where a generic procedure makes use of distribution-specific
iteratively weighted least squares approximations to the full
conditionals. The framework of distributional regression encompasses
many special cases relevant for treating nonstandard response
\mbox{structures} such as highly skewed nonnegative responses, overdispersed
and zero-inflated counts or shares including the possibility for
zero- and one-inflation.
We discuss distributional regression along a study
on determinants of labour incomes for full-time working males in
Germany with a particular focus on regional differences after the
German reunification. Controlling for age, education, work experience
and local disparities, we estimate full conditional income
distributions allowing us to study various distributional quantities
such as moments, quantiles or inequality measures in a consistent
manner in one joint model. Detailed guidance on practical aspects of
model choice including the selection of several competing distributions
for labour incomes and the consideration of different covariate effects
on the income distribution complete the distributional regression
analysis. We find that next to a lower expected income, full-time
working men in East Germany also face a more unequal income
distribution than men in the West, ceteris paribus.
\end{abstract}

%
\end{frontmatter}

\section{Introduction}\label{intro}

The analysis of determinants of labour incomes has a long tradition in
economics, dating back at least to \citet{Mincer1974}. His classical
wage equation includes potential labour market experience as well as
years of education as the most important determinants of human capital
which then translates into expected income [\citet{Lemieux2006}].
Additional possible determinants include age, actually realised labour
market experience, gender, regional information concerning the
residence of employees, or area of employment. One considerable
restriction of most analyses conducted so far is their sole focus on
the expected income given covariates, that is, the conditional mean. In
some cases, distributions are required, for example, for inequality
decomposition or to account for incomplete information due to
truncation or censoring. Then, the (log-)normal distribution
[\citet{Greene2008}, Chapter~19, \citet{Morduch2002}] is often implicitly
considered
(again with regression effects only on the mean) or one reverts to
local analyses by means of quantile regression [\citet{Autor2008,Galvao2013}].
More flexible types of distributions have so far mostly been used to
describe income distributions on a highly aggregated level, normally
the national level [\citet{Kleiber1996}].

\begin{figure}

\includegraphics{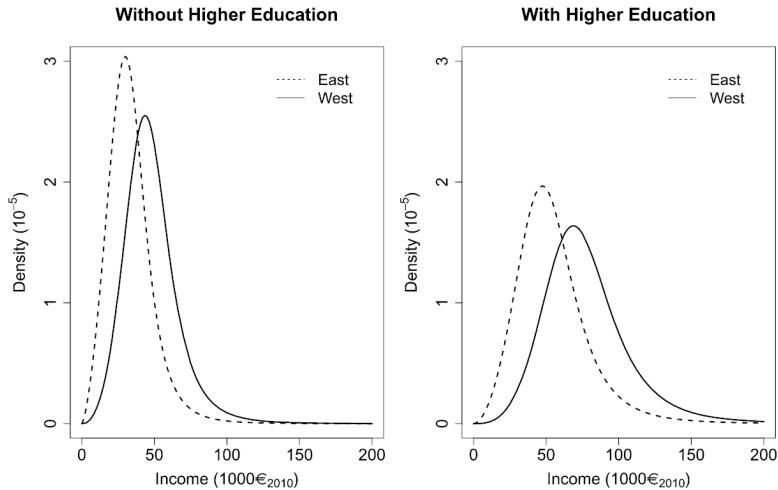}

\caption{SOEP data. Four conditional income distributions
for 42-year-old males with 19 years of working experience without
higher education (left) or with higher education (right) and living in
the East (dashed lines) or West (solid lines). Densities shown are
posterior means of the densities in our best model DA\_M1; compare
Section~\protect\ref{secmodelchoice} for details on the model specification.}
\label{Figcids}
\end{figure}

We utilise detailed, longitudinal information on incomes available from
the German socio-economic panel (SOEP) to derive a flexible, structured
additive distributional regression model for labour incomes of
full-time male workers. We consider several candidate distributions for
describing the nonnegative conditional income distributions, including
the log-normal distribution, the gamma distribution, the inverse
Gaussian distribution and the Dagum distribution. To obtain flexible
models, we allow for regression effects on potentially all parameters
of the income distribution, thereby overcoming the previous
concentration on expected incomes. As an illustration, consider the
income distributions visualised in Figure~\ref{Figcids} corresponding
to an ``average,'' full-time male worker with/without higher
education in East and West Germany. Here we find that the income
distributions differ considerably not only in terms of their
expectation but also with respect to other aspects of the distribution,
like the variance (see Section~\ref{secincome} for more details on the
analysis).

%

Some earlier attempts to define distributional regression models
comprise \citet{Biewen2005} or \citet{Donald2000}.
\citet{Biewen2005} suggest to decompose the population into a
coarse set of
subgroups for which parametric income distributions are estimated such
that the distributional form varies over the subgroups. \citet
{Donald2000} propose to vary location and scale parameters with
respect to covariates while the general shape of the distribution
remains fixed over the covariate set. Building on their work, we
propose to combine these approaches in the sense that conditional
income distributions are modelled parametrically as suggested
by \citet{Biewen2005}, while allowing for variation in the whole
distribution
(not just location and scale) with respect to covariates as specified
by \citet{Donald2000}.

Differences between East and West Germany have received considerable
attention in the economic literature [\citet{Biewen2000}, Fuchs-Sch{\"{u}}ndeln,\break Krueger and
Sommer (\citeyear{FuchsSchundeln2010}), \citet{Kohn2011}] and also consistently played
a major
role in the domestic political debate. Instead of solely taking a
macroeconomic perspective to look at income inequality in the East and
West at a highly aggregated level, we build a microeconomic foundation
to the analysis of income inequality. Thereby, we consider the effect
of various covariates on the conditional individual income distribution
underlying the aggregate income distribution. It is our hypothesis that
there are not only significant differences between East and West in the
conditional mean income but also in the conditional income inequality
aggravating the economic divide more than two decades after the reunification.

As a conceptual framework for our analyses, we extend the Bayesian
structured additive distributional regression models recently proposed
in \citet{KleKneLan2013} for zero-inflated and overdispersed
count data
regression to general types of univariate distributions. In this class
of regression models, all parameters of a potentially complex response
distribution are related to additive regression predictors in the
spirit of generalised additive models (GAMs). While the latter assume
responses to follow a distribution from the exponential family and
focus exclusively on relating the mean of a response variable to
covariates [see, e.g., \citet{RupWanCar2003,FahKneLan2004,fahkne13}, \citeauthor{Woo04a} (\citeyear{Woo04a,Woo08})], distributional
regression enables the consideration of basically any response
distribution and allows to specify regression predictors for all
parameters of this distribution. The main advantage of distributional
regression is that it provides a broad and generic framework for
regression models encompassing continuous, discrete and mixed
discrete-continuous response distributions and therefore considerably
expands the common exponential family framework.

Distributional regression is closely related to generalised additive
models for location, scale and shape (GAMLSS) as suggested by
\citet{RigSta2005}. We prefer the notion of distributional
regression for our
approach since in most cases, the parameters of the response
distribution are in fact not directly related to location, scale and
shape but are general parameters of the response distribution and only
indirectly determine location, scale and shape. For example, in case of
the Dagum distribution, there are three distributional parameters, but
none of them is directly related to a measure of location which is
jointly determined by all three parameters.

In GAMLSS, inference is commonly based on penalised maximum likelihood
estimation achieved via backfitting loops over the additive predictor
components.
In this paper, we consider a generic Bayesian treatment of
distributional regression relying on Markov chain Monte Carlo
simulation algorithms. To construct suitable proposal densities, we
follow the idea of iteratively weighted least squares proposals
[\citet{Gam1997,BreLan2006}] and use local quadratic
approximations to the
full conditionals in order to avoid manual tuning. Utilising explicit
derivations of the score function and expected Fisher information in
these approximations considerably enhances numerical stability as
compared to using numerical derivatives and the observed Fisher
information (which are frequently used in the R add-on package \texttt{gamlss} implementing penalised likelihood inference). The Bayesian
approach also has the advantage to provide credible intervals without
relying on asymptotic arguments. The full potential of distributional
regression is only exploited when the regression predictor is broadened
beyond the scope of simple linear or additive specifications. We will
consider structured additive predictors [\citet{BreLan2006,fahkne13}]
where each predictor is determined as an additive combination of
various types of functional effects, including nonlinear effects of
continuous covariates, spatial effects, random effects or varying
coefficient terms. 

Alternatives to distributional regression are provided by quantile and
expectile regression which also allow us to go beyond studying the mean
by focusing on local features of the response distribution, indexed by
a prespecified asymmetry parameter (the quantile or expectile level);
see \citet{KoeBas1978,NewPow1987} for the original references and
\citet{Koe2005,YuMoy2001,SchEil2009,SobKne2012} for more recent overviews.
Single quantiles or expectiles are elicitable [\citet{OsbRei1985,Gne2011b}] by considering asymmetrically weighted loss
functions and consistent estimates can be obtained under rather mild
conditions on the conditional distribution of the responses (basically
reducing to independence and the correct specification of the quantity
of interest). However, when interest focuses on the complete
conditional distribution or if distributional quantities such as the
Gini coefficient for inequality that are not elicitable by specifying a
corresponding loss function are desired, the direct specification of
distributional regression turns out to be advantageous. 

Many of the aspects discussed in the remainder of this paper (such as
choice of a suitable response distribution and adequate predictor
specifications, Bayesian inference, interpretation of estimation
results) are relevant beyond our application. We therefore provide an
analysis on the proportion of farm outputs achieved by cereals in the
application Supplement~A [\citet{suppa}, Section~A.2], to this paper as a
second example on distributional regression.

The remainder of the paper is structured as follows: Section~\ref{secsadr} provides a detailed introduction to distributional
regression and Bayesian inference along our case study on labour
incomes. Model choice concerning the type of the response distribution
and the specification of the regression predictors is treated in
Section~\ref{secmodelchoice}. Given the selected models, Section~\ref{secincome} provides empirical results on the regional disparities of
conditional incomes in East and West Germany. Additional material on
the application is provided in the application supplement Section~A.1. Section~\ref{secsummary} provides a summary and
comments on directions for future research. Finally, we summarise
general aspects of distributional regression with other types of
responses in the methodological Supplement~B [\citet{suppb}] which also comprises
details on Bayesian inference, derivations of required quantities for
the iteratively weighted least squares proposals and simulation studies.

\section{Distributional regression}\label{secsadr}

As a conceptual framework for our analysis of labour incomes and their
regional disparities, we consider distributional regression models
where, conditional on all available covariate information summarised in
the vector $\bolds{\nu}_i$, the response variables $y_1,\ldots,y_n$ are
assumed to be independently distributed with $K$-parametric densities
$p(y_i|\vartheta_{i1},\ldots,\vartheta_{iK})\equiv p_i$. The
conditional distribution $p_i$ of observation $y_i$ given $\bolds{\nu
}_i$ is
indexed by the (in general covariate-dependent) distributional
parameters $\vartheta_{i1},\ldots,\vartheta_{iK}$. Each parameter
$\vartheta_{ik}$, $k=1,\ldots,K$ is then related to a semiparametric,
additive predictor $\eta_{i}^{\vartheta_k}$ defined in terms of
(potentially different) subvectors of the covariate vector $\bolds{\nu}_i$.
Similarly, as in generalised linear models, a suitable (one-to-one)
response function is utilised to map the predictor to the parameter of
interest, that is, $\vartheta_{ik}=h^{\vartheta_k}(\eta
_{i}^{\vartheta
_k})$, where the superscript $\vartheta_k$ in the predictors and
response functions indicates that we are dealing with $K$ predictors
specific to the different distributional parameters instead of only one
single predictor as in mean regression. The response function is chosen
to ensure appropriate restrictions on the parameter space such as the
exponential function $\vartheta_{ik}=\exp(\eta_{i}^{\vartheta_k})$ to
ensure positivity. We discuss specific choices for distributional
regression of labour incomes after having introduced our data in more detail.

\subsection{German labour income data}

For studying conditional income distributions in Germany, we utilise
information from the German Socio-Economic Panel [\citet{Wagner2007}].
More specifically, we consider real gross annual personal labour income
in Germany as defined in \citet{Bach2009} for the years 2001 to 2010.
We deflate the incomes by the consumer price index [\citet
{StatistischesBundesamt2012b}], setting 2010 as our base year. Thus,
all incomes are expressed in real-valued 2010 Euros from here on.

Following the standard literature, we only look at the income of males
in full-time employment [see, among others, \citet
{Dustmann2009,Card2013}] in the age range 20--60. This yielded 7216
individuals for
whom we considered the income trajectories from the ten year period.
For each individual, we used every observation for which all required
dependent and independent variables were available, yielding a total of
$n=40{,}965$ observations. Naturally, this implies that for some
individuals we do not have full longitudinal coverage over the whole
ten year period.

As covariates, we consider educational level measured as a binary
indicator for completed higher education (according to the UNESCO
International Standard Classification of Education 1997 provided in the
SOEP) in effect coding ($\mathit{educ}$), age in years ($\mathit{age}$), previous labour market experience in years ($\mathit{lmexp}$),
the calendar time ($t$), information on the geographical district
(\textit{Raumordnungsregion}) representing the area of residence
($s$) and a binary indicator in effect coding for districts
belonging to the eastern part of Germany ($\mathit{east}$). A
description of the data set is given in Table~A1;
details on the specifications for the different effect types will be
provided in Section~\ref{subsecSTAR}.

A common assumption in economic analyses of income is that incomes
$y_i$ are log-normally distributed with covariate-dependent location
parameter $\eta_i$ (corresponding to the mean of the log-transformed
incomes) and a constant scale parameter $\sigma^2$. For an observation
$i$ collected at time point $t_i$, a suitable semiparametric predictor
(dropping the dependence on the parameter $\vartheta_k$) could then be
specified as
%
\begin{eqnarray}
\eta_{i} &=& \beta_{0} + \mathit{educ}_{i}
\beta_{1} + f_{1}(\mathit{age}_{i})+
\mathit{educ}_{i} f_{2}(\mathit{age}_{i})
\nonumber
\\[-8pt]
\label{eqpredictor}
\\[-8pt]
\nonumber
&&{}+
f_{3}(\mathit {lmexp}_{i}) + f_{\mathrm{spat}}(s_i)+f_{\mathrm{time}}(t_i),
\end{eqnarray}
where $\beta_0$ represents the overall intercept, $\beta_{1}$ captures
the effect of higher education, $f_{1}(\mathit{age})$ and
$f_{2}(\mathit
{age})$ are nonlinear effects of age capturing also the interaction
with the educational status, $f_{3}(\mathit{lmexp})$ is the nonlinear
effect of previous labour market experience, $f_{\mathrm{spat}}(s)$ is a spatial
effect capturing heterogeneity at the level of the districts $s$, and
$f_{\mathrm{time}}(t)$ is an effect specific for the calendar year $t$. In a
second step, the spatial effect can further be decomposed into
%
\begin{equation}
\label{eqpredictor2}
f_{\mathrm{spat}}(s) = \mathit{east}_s
\gamma_1 + g_{\mathrm
{str}}(s) + g_{\mathrm{unstr}}(s),
\end{equation}
where $\gamma_1$ captures the difference between the eastern and
western part of Germany and $g_{\mathrm{str}}(s)$ and $g_{\mathrm
{unstr}}(s)$ represent spatially structured (smooth) or unstructured
(unsmooth) district-specific effects. Note that, in addition to the
East--West indicator $\mathit{east}$, more district-specific information
could be included if desired. While this decomposition could simply be
plugged into (\ref{eqpredictor}) to obtain a reduced-form
specification, it can also be interpreted as a hierarchical multilevel
specification where we differentiate between an individual-specific
level in (\ref{eqpredictor}) and a region-specific level in (\ref{eqpredictor2}). Further details on the predictors and associated
priors will be discussed in Section~\ref{subsecSTAR}.

\subsection{Potential response distributions}\label{secpotential-response-distributions}

One of the great advantages of structured additive distributional
regression is the wide range of distribution types that can be
modelled. Since labour income is by definition positive, we will
restrict ourselves to four nonnegative distributions summarised in
Table~\ref{tabpotresp}. For a more comprehensive list of distributions
supported by the distributional regression framework, see Section~B.1.1.

As noted, the standard conditional distribution type in econometric
income analyses is the log-normal distribution. Next to its theoretical
appeal from an economic perspective [see \citet{Arnold2008}, page 122],
it has the advantage that it makes the vast statistical inference
machinery built around Gaussian regression available to researchers.
However, \citet{Atkinson1975} and others have noted that, at
least for
the aggregate income distribution, the log-normal distribution fit is
problematic at times, especially for the upper tail of the distribution.

Partly as a consequence, various other distribution types have thus
been suggested for the modelling of income distributions. \citet
{Salem1974} proposed the gamma distribution as a suitable alternative
to the log-normal distribution. One of its advantages is that its
estimation is possible within the framework of generalised linear
models as the distribution belongs to the exponential family (as long
as covariate effects are restricted to the mean).

The third distribution we consider also belongs to the exponential
family (if the second parameter is assumed to be independent of
covariates). The inverse Gaussian distribution has to our knowledge not
been used in the context of modelling income distributions yet. But for
other nonnegative distributions with a similar economic rationale,
like the distribution of claim sizes arising in car
insurance~[Heller, Stasinopoulos and Rigby (\citeyear{HelStaRig2006}), \citet{KleDenKneLan2013}], it has shown
to perform well due to
its flexibility in modelling extreme right skewness. As it is
conceivable that some conditional income distributions also portray
such extreme skewness, we decided to also consider this distribution type.

%
\begin{table}
\tabcolsep=0pt
\caption{Selected candidate distributions; for a more
comprehensive list see Table~\textup{B1}}
\label{tabpotresp}
\fontsize{7.5pt}{10pt}\selectfont{\begin{tabular*}{\tablewidth}{@{\extracolsep{\fill}}lccc@{}}
\hline
\textbf{Name} & \textbf{Density} & \textbf{Parameters} & \textbf
{Response functions}\\
\hline
Log-normal & $p(y|\mu,\sigma^2)=\frac{1}{\sqrt{2\pi\sigma^2}
y}\exp
(-\frac{(\log(y)-\mu)^2}{2\sigma^2} )$ &$\mu\in\mathbb
{R},\sigma
^2>0$ &
$h^{\mu}(\eta)=\eta,h^{\sigma^2}(\eta)=\exp(\eta)$\\[3pt]
Inverse Gaussian & $p(y|\mu,\sigma^2)=\frac{1}{\sqrt{2\pi\sigma^2}
y^{3/2}}\exp (-\frac{(y-\mu)^2}{2 y\mu^2\sigma^2}
)$&$\mu
,\sigma^2>0$ & $h^{\mu}(\eta)=h^{\sigma^2}(\eta)=\exp(\eta)$ \\[3pt]
Gamma & $p(y|\mu,\sigma)= (\frac{\sigma}{\mu} )^{\sigma
}\frac
{y^{\sigma-1}}{\Gamma(\sigma)}\exp (-\frac{\sigma}{\mu
}y
)$&$\mu,\sigma>0$& $h^{\mu}(\eta)=h^{\sigma}(\eta)=\exp(\eta)$
\\[3pt]
Dagum & $p(y|a,b,c)=\frac{a c y^{a c-1}}{b^{a c} (1+
(y/b
)^{a} )^{c+1}}$&$a,b,c>0$& $h^{a}(\eta)=h^{b}(\eta)=h^{c}(\eta
)=\exp(\eta)$ \\
\hline
\end{tabular*}}
\end{table}

The last distribution we consider is the Dagum distribution
[\citet{Dagum1977}] which belongs to the beta-type size
distributions that
have seen considerable attention in the literature on modelling
(aggregate) income distributions [see \citet{Kleiber2003}]. One
of its
appealing properties is that towards the upper end of the distribution
its shape mirrors the one of the Pareto distribution which is generally
assumed to provide a good approximation for the income distribution for
the top percentiles of the (aggregate) income distribution [\citet
{Piketty2007}].


\subsection{Structured additive predictors and associated
priors}\label{subsecSTAR}

\subsubsection*{Generic representation}
While considering a specific instance of a structured additive
predictor for the analysis of income, a generic structured additive
predictor for parameter $\vartheta_{ik}$ is given by
%
\begin{equation}
\label{eqgenpred2} \eta_{i}^{\vartheta_k} = \beta_0^{\vartheta_k}
+ f_{1}^{\vartheta
_k}(\bolds{\nu}_i) + \cdots+
f_{J_k}^{\vartheta_k}(\bolds{\nu}_i),
\end{equation}
where $\beta_0$ represents the overall level of the predictor and the
functions $f_j^{\vartheta_k}(\bolds{\nu}_i)$, $j=1,\ldots,J_k$,
relate to
different covariate effects defined in terms of the complete covariate
vector $\bolds{\nu}_i$. Note that each distribution parameter 
may depend on different covariates and a different number of effects
$J_k$, but we suppress this possibility (as well as the parameter
index) in the following.

In structured additive regression, each function $f_j$ is approximated
by a linear combination of $D_j$ appropriate basis functions, that is,
\[
f_j(\bolds{\nu}_i) = \sum
_{d_j=1}^{D_j}\beta _{j,d_j}B_{j,d_j}(
\bolds{\nu}_i)
\]
such that in matrix notation we can write $\mathbf{f}_j=(f_j(\bolds
{\nu}
_1),\ldots
,f_j(\bolds{\nu}_n))'=\mathbf{Z}_j\bolds{\beta}_j$, where $\mathbf{Z}
_j[i,d_j]=B_{j,d_j}(\bolds{\nu}
_i)$ is a design matrix and $\bolds{\beta}_j$ is the vector of coefficients
to be estimated.
To ensure identifiability specific constraints
representing for example centring of the functional effects are added,
see Section~B.2.2 for further details.
The basis function representation then leads to the
following matrix representation of the generic predictor (\ref{eqgenpred2}):
%
\begin{equation}
\label{eqgenpred}
\bolds{\eta}= \beta_0\mathbf{1}+\mathbf{Z}_1
\bolds{\beta }_1+\cdots+\mathbf{Z}_{J}\bolds{
\beta}_{J}.
\end{equation}
For each of the parameter vectors $\bolds{\beta}_j$ we can then either
assume a hierarchical specification, where $\bolds{\beta}_j$ is
related to
another structured additive predictor (as in the case of the spatial
effect in our example), or we directly assume the multivariate normal prior
%
\begin{equation}
\label{eqpriorspec}
p\bigl(\bolds{\beta}_j|\tau_j^2
\bigr) \propto \biggl(\frac{1}{\tau
_j^2} \biggr)^{({\mathrm{rk}(\mathbf{K}_j)})/{2}}\exp \biggl(-
\frac{1}{2\tau
_j^2}\bolds{\beta}_j'\mathbf{K}
_j\bolds{\beta} _j \biggr)
\end{equation}
with\vspace*{1pt} (potentially rank-deficient) precision matrix $\mathbf{K}_j$ and prior
smoothing variance~$\tau_j^2$. The latter\vspace*{1pt} is assigned an inverse gamma
hyperprior $\tau_j^2~\sim\operatorname{IG}(a_j,b_j)$ (with
$a_j=b_j=0.001$ as a
default option) in order to obtain a data-driven amount of smoothness.

A detailed discussion of terms that fit into the generic predictor
framework (in the context of mean regression) is provided in
\citet{FahKneLan2004} and \citet{fahkne13}, Chapters~8~and~9.

In the following, we will discuss suitable specifications and prior
assumptions for the hierarchical predictor defined in (\ref
{eqpredictor}) and (\ref{eqpredictor2}). Note that we drop the
dependence on the distributional parameter indicated by the superscript
$\vartheta_k$, the observation index $i$ and the function index $j$ to
simplify notation. Hierarchical extensions are treated in detail in
\citet{LanUmlWecHarKne2012}.

\subsubsection*{Linear effects} For all parametric, linear effects, we
assume a flat, noninformative prior. This may be considered the
limiting case of a multivariate Gaussian prior with high dispersion
which can also be used to achieve regularisation in the case of
high-dimensional parameter vectors. In our analyses, we assume linear
effects for the intercept and the educational indicator, as well as for
the East--West indicator.

\subsubsection*{Continuous covariates} For the effects of age and previous
work experience, assuming a linear effect is probably too restrictive.
We therefore consider P(enalised)-splines [\citet{eilers}] as a flexible
device for including potentially nonlinear effects $f(x)$ of a
continuous covariate $x$. In a first step, $f(x)$ is approximated by a
linear combination of $D$ B-spline basis functions $B_d(x)$ that are
constructed from piecewise polynomials of a certain degree $l$ upon an
equidistant grid of knots, $f(x) = \sum_{d=1}^D\beta_dB_d(x)$. To avoid
the requirement of choosing an optimal number of knots together with
optimal knot positions, \citet{eilers} regularise the function estimate
by augmenting a difference penalty to the fit criterion. In our
Bayesian framework, the stochastic analogue is to assume a first or
second order random walk
\begin{eqnarray*}
\beta_{d}&=&\beta_{d-1}+\varepsilon_d,\qquad
 d=2,\ldots,D,
\\
\beta_{d}&=&2\beta_{d-1}-\beta_{d-2}+
\varepsilon_d,\qquad d=3,\ldots,D
\end{eqnarray*}
with Gaussian errors $\varepsilon_d\sim\mathrm{N}(0,\tau^2)$ and
noninformative priors for $\beta_{1}$ or $\beta_{1}$ and~$\beta_{2}$
[\citet{LanBre2004}]. The joint prior of all basis coefficients
$\bolds{\beta}
=(\beta_1,\ldots,\beta_D)'$ can then be shown to be a (partially
improper) multivariate Gaussian distribution with zero mean and
precision matrix $\mathbf{K}=\mathbf{D}'\mathbf{D}$, where $\mathbf
{D}$ is a difference matrix of
appropriate order. In our analysis, we use twenty inner knots, a cubic
spline basis and a second order random walk prior as the default
specification for penalised splines.

In the case of the age effect, we allow for separate functions for
individuals with high and low levels of education. This is achieved by
the inclusion of the varying coefficient term [\citet{HasTib1993}]
$f_{2}(\mathit{age})$ such that the age effect is given by
$f_1(\mathit
{age})-f_2(\mathit{age})$ for individuals with low educational level
and $f_1(\mathit{age})+f_2(\mathit{age})$ for individuals with high
educational level. In this case, a~penalised spline can be assumed for
function $f_2(\mathit{age})$ as well.

\subsubsection*{Random effects} Penalised splines can in principle also be
considered to represent the temporal effect $f_{\mathrm{time}}(t)$ in (\ref
{eqpredictor}). However, since in economic research temporal effects
such as ours are generally considered by year-specific effects, we do
not impose the smoothness assumption implied by penalised splines.
We therefore consider a random effects specification where separate
regression effects $\beta_{t}=f_{\mathrm{time}}(t)$ are assumed for the distinct
time points. An i.i.d.~Gaussian prior with random effects variance
$\tau
^2$ is then placed on the coefficients $\bolds{\beta}=(\beta
_1,\ldots
,\beta
_T)'$. Similarly, random effects priors can be used for any other
grouping variable with levels $\{1,\ldots,G\}$ present in the data.

Note that we have not included individual-specific random effects. The
reason for this is that we are specifically interested in the
unobserved heterogeneity among individuals with similar covariate sets
which finds expression in income inequality among them. In some sense
our analysis is thus systematically different from standard regression
techniques which pursue to eradicate the stochastic component or at
least reduce it to a minimum. The inclusion of individual-specific
effects goes a long way towards seemingly achieving this aim, as the
share of the variance left to the error term is drastically reduced.
However, the inferential gain obtained thereby could be expressed as
follows: including individual-specific effects, we have found that
incomes are largely different because individuals are different. While
there are some analyses where such eradication of variance is useful,
it sheds little insights on the nature of inequality at the
disaggregated level since we are unable to disentangle the differences
between individuals in a meaningful way.

\subsubsection*{Spatial effects} For the spatial effect $f_{\mathrm{spat}}(s)$
defined upon the discrete, spatial variable $s\in\lbrace1,\ldots,
S\rbrace$ which denotes the different regions in the data set, we
assume a hierarchical predictor specification following \citet
{LanUmlWecHarKne2012}. In fact, equation (\ref{eqpredictor2}) merely
defines a second structured additive predictor where now the distinct
spatial regions define the unit of observation. As a consequence, any
type of regression effect that is specific for the region can be
included on this level. In our case, the East--West indicator is one
such example that is assigned a parametric effect with flat prior.

In addition, we consider the spatially structured and spatially
unstructured effects $g_{\mathrm{str}}(s)$ and $g_{\mathrm{unstr}}(s)$, respectively. In
both cases, separate regression effects $\beta_{\mathrm{str},s}=g_{\mathrm{str}}(s)$ and
$\beta_{\mathrm{unstr},s}=g_{\mathrm{unstr}}(s)$ are assumed for each of the regions, but
the effects differ in terms of their prior assumptions. For the
structured spatial effect, we assume spatial correlations defined
implicitly by assuming a Gaussian Markov random field prior [\citet
{RueHel2005}] for a suitable neighbourhood structure derived from the
spatial orientation of the data. The most common case would be to treat
two regions as neighbours if they share a common boundary. If $\partial
_s$ denotes the set of all neighbours of region $s$, the Markov random
field prior then assumes
%
\begin{equation}
\label{secgmrf} \beta_{\mathrm{str},s}|\beta_{\mathrm{str},r}, r\neq s,
\tau^2\sim\mathrm{N} \biggl(\sum_{r\in
\partial_s}
\frac{1}{N_s}\beta_{\mathrm{str},r},\frac{\tau^2}{N_s} \biggr)
\end{equation}
with number of neighbours of region $s$ denoted as $N_s$. Consequently,
the conditional mean of $\beta_{\mathrm{str},s}$ given all other coefficients is
the average of the neighbouring regions. It can be shown that the
conditional normal distributions specified in (\ref{secgmrf})
correspond to a multivariate, partially improper normal distribution
with zero mean and precision matrix given by the adjacency matrix
induced by the neighbourhood structure.

For the unstructured spatial effect, we consider an i.i.d. Gaussian
prior, that is, we assume a random effects prior specification. The
rationale for considering both a structured and unstructured part of
the spatial effect is that they are surrogates for unobserved spatial
heterogeneity which may either be spatially structured (i.e., spatially
smooth) or unstructured.

\subsection{Bayesian inference}\label{secinference}

To perform Bayesian inference, we consider\break Markov chain Monte Carlo
(MCMC) simulation techniques and develop suitable proposal densities
based on iteratively weighted least squares (IWLS) approximations to
the full conditionals. The derivation of the approximations and the
complete algorithm are documented in Section~B.2. Here, we only sketch the essential parts.

\subsubsection*{IWLS proposals for regression coefficients} The regression
coefficients $\bolds{\beta}_j$ are proposed from $\mathrm{N}(\bolds
{\mu}_j,\mathbf{P}_j^{-1})$
with expectation and precision matrix
\[
\bolds{\mu}_j = \mathbf{P}_j^{-1}
\mathbf{Z}'_j \mathbf{W}(\mathbf {z}-\bolds{
\eta}_{-j}), \qquad\mathbf{P}_j = \mathbf{Z}
'_j \mathbf{W}\mathbf{Z}_j+
\frac{1}{\tau_j^2}\mathbf{K}_j,
\]
where $\mathbf{W}$ is a diagonal matrix of working weights
$w_{i}=\mathbb{E}
(-\partial^2 l/\partial\eta_i^2)$, $\mathbf{z}=\bolds{\eta}+
(\mathbf{W})^{-1}\mathbf{v}$ is a working response depending on the score vector
$\mathbf{v}
=\partial l/\partial\bolds{\eta}$ and $\bolds{\eta}_{-j}=\bolds
{\eta}-\mathbf{Z}
_j\bolds{\beta}
_j$ is the predictor without the $j$th component. The working weights
and the score vector are specific for the chosen response distribution
and induce an automatic adaptation to the form of the full conditional
without requiring manual tuning.

\subsubsection*{Updates for the smoothing variances}
The smoothing variances $\tau_j^2$ can be sampled in a Gibbs update
where $\tau_j^2|\cdot\sim\operatorname{IG}(a'_j,b'_j)$, with
updated parameters
$a'_j=\frac{\mathrm{rk}(\mathbf{K}_{j})}{2}+a_j$, $b'_j=\frac
{1}{2}\bolds{\beta}'_j
\mathbf{K}
_{j}\bolds{\beta}_j+b_j$.

\subsubsection*{Working weights} The specification of the working weights
$\mathbf{W}$ involves the expectations of the negative second
derivatives of
the log-likelihood which improved both mixing and acceptance rate in
comparison with the (seemingly simpler) approach of using the negative
second derivative without deriving the expectation. Furthermore,
invertibility of the precision matrix $\mathbf{P}_j$ is ensured for many
distributions when using the expectation since the working weights are
then nonnegative. Explicit derivations for both the distributions
utilised for analysing labour incomes and the additional distributions
summarised in Table~B1 can be found in Section~B.2.3.

\subsubsection*{Propriety of the posterior} Propriety of the posterior in
distributional regression can be ensured when combining the assumptions
considered in \citet{KleKneLan2013} for count data regression with
appropriate restrictions on the densities. These need to be bounded or
integrable with respect to the predictors, whereby at least one
observation fulfilling the latter assumption is required. Note that
integrability of the densities can be assured by the assumption that
none of the distributional parameters is on the boundary of the
parameter space (an assumption that would also have to be made to apply
standard maximum likelihood asymptotics).

\subsubsection*{Software} Our Bayesian approach to distributional
regression is implemented in the free, open source software BayesX
[\citet{BelBreKleKneLanUml2015}]. As described in \citet
{LanUmlWecHarKne2012}, the implementation makes use of efficient
storing mechanisms for large data sets and sparse matrix algorithms for
sampling from multivariate Gaussian distributions. An R interface to
BayesX is provided in the R add-on package \texttt{bamlss} [\citet
{UmlKleLanZei2014}].

\subsubsection*{Empirical evaluation}\label{secempirical-evaluation} We
compared the empirical performance of the proposed Bayesian approach to
the frequentist GAMLSS framework in two simulation scenarios and also
investigated the performance of the deviance information
criterion~[DIC, \citet{SpiBesCarLin2002}] for choosing between competing
models. The studies and their outcomes are documented in more detail in
Section~B.3. A summary on the ability of the DIC for
model choice is given in Section~\ref{secmodelchoice} and for the
comparison with the frequentist approach (denoted as ML) in the following:
\begin{longlist}[(2)]
\item[(1)] \emph{Comparison with ML in additive models}. In purely additive
models, the point estimates and corresponding posterior means, as well
as their mean squared errors (MSEs), are very similar. However,
coverage rates based on asymptotic maximum likelihood theory for ML are
far too narrow in several distribution parameters. In particular, for
the Dagum distribution, rates for all three parameters are far from the
desired coverage level, while the credible intervals of the Bayesian
approach are still reliable (albeit being usually slightly too
conservative); compare, for example,~Figure~B2.

\item[(2)]  \emph{Comparison with ML in geoadditive models}.  10\% of the
estimation runs of ML failed before convergence. MSEs of the spatial
effect (based on a Markov random field) are slightly smaller for the
Bayesian approach compared to ML. While the MSEs of the other effects
do not deteriorate for our proposed method, we observe partly
increasing MSEs for ML.
\end{longlist}

\section{Model choice}\label{secmodelchoice}

In any application of distributional regression, one faces important
model choice decisions: choosing the most appropriate out of a set of
potential response distributions and selecting adequate predictor
specifications for each parameter of these distributions. For our
application on conditional income distributions, we consider the
inverse Gaussian (IG), log-normal (LN), gamma (GA) and Dagum (DA)
distribution as candidate distributions. A general predictor that could
now be utilised for any of the parameters of these distributions was
already introduced in equations~(\ref{eqpredictor}) and (\ref
{eqpredictor2}). Instead of performing a complete stepwise model
selection for each distribution, we study the following model specifications:
\begin{longlist}[(M1)]
\item[(M1)] All distributional parameters are related to a predictor of type
(\ref{eqpredictor}). For the spatial effect, we only include the
unstructured effect since it turned out in exploratory analyses that
the smooth component has only negligible impact.
\item[(M2)] Instead of modelling all parameters in terms of covariates, the
model structure of M1 is only applied to the parameters $\mu$ in the
case of LN, IG and GA, and $b$ in case of DA. The parameters $a, c,
\sigma,\sigma^2$ are considered to be equal across all individuals.
This corresponds to a usual GAM specification with focus on conditional means.
\item[(M3)] All parameters are modelled in analogy to M1 except that the
random effect for calendar time and the complete spatial effect
(including the East--West indicator) are not included in the parameters
$a, c, \sigma,\sigma^2$.
\end{longlist}
In total, we therefore end up with 12 models to compare. In the
following, we will discuss different options for conducting this
comparison and will also comment on their wider applicability in the
context of model choice for distributional regression.


\subsection{Deviance information criterion}
The deviance information criterion  (DIC) is a commonly used criterion
for model choice in Bayesian inference that has become quite popular\vspace*{1pt}
due to the fact that it can easily be computed from the MCMC output. If
${\bolds{\theta}}^{[1]},\ldots,{\bolds{\theta}}^{[T]}$ is a MCMC
sample from the
posterior for the complete parameter\vspace*{1pt} vector ${\bolds{\theta}}$, the
DIC is
given by $\overline{D({\bolds{\theta}})}+ \mathit{pd}=2 \overline
{D({\bolds{\theta}})}-D(\overline{{\bolds{\theta}}})=\frac
{2}{T}\sum D({\bolds{\theta}}
^{[t]})-D(\frac
{1}{T}\sum{\bolds{\theta}}^{[t]})$, where $D({\bolds{\theta
}})=-2\log(f(\mathbf{y}
|{\bolds{\theta}}))$ is the model deviance and $\mathit
{pd}=\overline
{D({\bolds{\theta}})}-D(\overline{\bolds{\theta}})$ is an
effective parameter count.

The DIC can be used to discriminate between types of response
distributions as well as different predictor specifications for a fixed
distribution. The latter can also be implemented in a stepwise model
choice strategy. However, since the DIC is sample-based, small
differences of DIC values for competing models may induce a region of
indecisiveness. If in such a situation sparser models are desired, the
DIC-based selection of covariate effects can be assisted by only
including significant effects, that is,~effects for which the credible
interval of a certain level does not contain the zero (parametric
effects) or the zero line (nonparametric effects); compare also
Section~B.3.3.2.

For distributions and models considered in our applications, we
conducted simulations on the performance of the DIC which are
documented in detail in Section~B.3.3. The basic
outcome is that the DIC can discriminate between competing response
distributions although differences can be rather small depending on
what distributions are compared. Concerning the identification of
relevant covariates, we focused on spatial effects and found that the
DIC usually is in clear favour of the true model if a relevant effect
is omitted. In the reverse situation, that is,~irrelevant information
is included, the DICs of the true models are only slightly smaller, but
then the irrelevant covariate mainly yields an insignificant effect
(i.e.,~the 95\% credible interval of each region contains zero) and
would thus be excluded under the aim of a sparser model. For count data
distributional regression models, the performance of the DIC was also
positively evaluated by \citet{KleKneLan2013} who compare several
misspecified models to the true model in terms of the DIC.

The DIC values for the 12 income regression models under consideration
are documented in Table~\ref{tabdic-scores-income} and indicate a
clear preference for the model DA\_M1. In general, it is noticeable
that the DIC favours our flexible model specifications (M1) compared to
the simplified versions (M2, M3).

\begin{table}[b]
\caption{Comparison of DIC values (calculated based on
the complete data set) and average scores obtained from ten-fold
cross-validation}\label{tabdic-scores-income}
\begin{tabular*}{\tablewidth}{@{\extracolsep{\fill}}lccccc@{}}
\hline
\textbf{Distribution}&\textbf{DIC}&
\textbf{Quadratic score} &
\textbf{Logarithmic score}&
\textbf{Spherical score} & \textbf{CRPS}\\
\hline
LN\_M1 & 179{,}090 & 0.130 & $-$2.436 & 0.362 & $-$2.158\\
LN\_M2 & 180{,}533 & 0.126 & $-$2.460 & 0.357 & $-$2.141\\
LN\_M3 & 179{,}451 & 0.130 & $-$2.435 & 0.362 & $-$2.163\\[3pt]
IG\_M1 & 184{,}614 & 0.146 & $-$2.274 & 0.378 & $-$1.620 \\
IG\_M2 & 189{,}702 & 0.138 & $-$2.314 & 0.366 & $-$1.677 \\
IG\_M3 & 186{,}494 & 0.144 & $-$2.282 & 0.374 & $-$1.642\\[3pt]
GA\_M1 & 177{,}453 & 0.161 & $-$2.172 & 0.396 & $-$1.274 \\
GA\_M2 & 178{,}736 & 0.156 & $-$2.181 & 0.392 & $-$1.279\\
GA\_M3 & 177{,}971 & 0.160 & $-$2.174 & 0.395 & $-$1.277\\[3pt]
DA\_M1 & \textbf{172{,}421} & \textbf{0.168} & \textbf{$-$2.103} &
\textbf{0.405} & \textbf{$-$1.266}\\
DA\_M2 & 173{,}791 & 0.164 & $-$2.120 & 0.402 & $-$1.274\\
DA\_M3 & 172{,}790 & 0.167 & $-$2.108 & 0.404 & $-$1.270 \\
\hline
\end{tabular*}
\end{table}

\subsection{Quantile residuals}

For continuous random variables, it is a well-known result that the
cumulative distribution function $F(\cdot)$ evaluated at the random
variable $y_i$ yields a uniform distribution on $[0,1]$. As a
consequence, quantile residuals defined as $\hat r_i=\Phi
^{-1}(F(y_i|\hat{\bolds{\vartheta}}_i))$, with the inverse cumulative
distribution function (c.d.f.) of a standard normal distribution $\Phi
^{-1}$ and $F(\cdot|\hat{\bolds{\vartheta}}_i)$ denoting c.d.f. with estimated
parameters $\hat{\bolds{\vartheta}}_i=(\hat\vartheta_{i1},\ldots
,\hat
\vartheta
_{iK})'$ plugged in, should at least approximately be standard normally
distributed if the correct model has been specified [\citet
{DunSmy1996}]. In practice, the residuals can be assessed graphically
in terms of quantile--quantile-plots: the closer the residuals are to
the bisecting line, the better the fit to the data. We suggest to use
quantile residuals as an effective tool for deciding between different
distributional options where strong deviations from the bisecting line
allow us to sort out distributions that do not fit the data well.

Quantile residuals are closely related to the probability integral
transform (PIT) which considers $u_i=F(y_i|\hat{\bolds{\vartheta
}}_i)$ without
applying the inverse standard normal c.d.f. If the estimated model is a
good approximation to the true data generating process, the $u_i$ will
then approximately follow a uniform distribution on $[0,1]$. As a
graphical device, histograms of the $u_i$ are then typically considered.

\begin{figure}[b]

\includegraphics{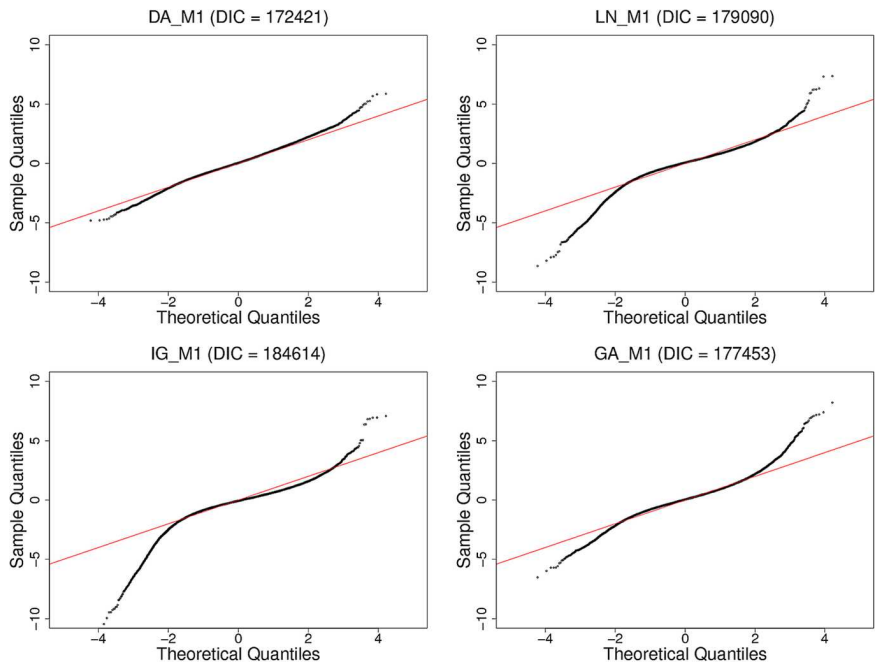}

\caption{Comparison of quantile
residuals for the full models DA\_M1 (topleft), LN\_M1 (topright), IG\_
M1 (bottomleft),
GA\_M1 (bottomright).}\label{figresincomeM1}
\end{figure}

Quantile residual plots for the models of type M1 are shown in
Figure~\ref{figresincomeM1}. Similar outcomes for model types M2/M3
and PITs for the models M1 can be found in Figures~A1 and~A2, respectively. We prefer
quantile residuals in the quantile--quantile-plot representation since
they avoid the requirement to define breakpoints in the construction of
the histogram.

While none of the distributions provides a perfect fit for the data,
the Dagum distribution turns out to be most appropriate for residuals
in the range between $-2$ and $2$ but deviates from the diagonal line
for extreme residuals. In contrast, the log-normal and inverse Gaussian
distribution seem to have problems in capturing the overall shape of
the income distribution, resulting in sigmoidal deviations from the
diagonal. Residuals of the gamma model are reasonable in the range
between $-2$ and $2$ (similar to the Dagum distribution) but deviate
more strongly from the diagonal for extreme residuals.

\subsection{Proper scoring rules}

Gneiting and Raftery (\citeyear{GneRaf2007}) propose proper scoring rules as summary
measures for
the evaluation of probabilistic forecasts, that is,~to evaluate the
predictive ability of a statistical model. We consider three common
scores, namely, the Brier or quadratic score (QS), the logarithmic
score (LS) and the spherical score (SPS). For continuous response
distributions with density $p_r(y)=p(y|\vartheta_{r1},\ldots
,\vartheta
_{rK})$ and a given new realisation $y_{\mathrm{new}}$, these are
defined as
\begin{eqnarray*}
\mathrm{LS}(p_r,y_{\mathrm{new}}) &=& \log
\bigl(p_r(y_\mathrm{new})\bigr),
\\
\mathrm{SPS}(p_r,y_\mathrm{new}) &=& \frac{p_r(y_\mathrm
{new})}{
(\int|p_r(y)|^2\,\mathrm{d}y )^{1/2}},
\\
\mathrm{QS}(p_r,y_\mathrm{new}) &=& 2p_r(y_\mathrm{new})-
\int \bigl|p_r(y)\bigr|^2\,\mathrm{d}y.
\end{eqnarray*}
Appropriate definitions for discrete as well as mixed discrete
continuous responses are provided in Section~B.1.2.
As a fourth alternative, we consider the continuous ranked probability
score (CRPS)
\[
\mathrm{CRPS}(p_r,y_\mathrm{new}) = -\int
_{-\infty}^{\infty} \bigl(F_r(y)-
\mathbh{1}_{\lbrace y\geq y_\mathrm{new}\rbrace} \bigr)^2\,\mathrm{d}y,
\]
where $F_r$ is the cumulative distribution function corresponding to
the density $p_r$ [\citet{GneRan2011}]. \citet{LaiTam2007}
showed that
the CRPS score can also be written as
\[
\mathrm{CRPS}(p_r,y_\mathrm{new}) =-2\displaystyle\int
_{0}^{1} (\mathbh{1}_{\lbrace y_\mathrm{new}\leq F_r^{-1}(\alpha)\rbrace
}-\alpha )
\bigl(F_r^{-1}(\alpha)-y_\mathrm{new} \bigr)\,\mathrm
{d}\alpha,
\]
where $F_r^{-1}(\alpha)$ is the quantile function of $p_r$ evaluated at
the quantile level $\alpha\in(0,1)$. This formulation allows not only
to look at the sum of all score contributions (i.e., the whole
integral) but also to perform a quantile decomposition and to plot the
mean quantile scores versus $\alpha$ in order to compare fits of
specific quantiles [\citet{GneRan2011}]. This decomposition is
especially helpful in situations where the quantile score can be
interpreted as an economically relevant loss function [\citet{Gne2011}].

In practice, we obtain the probabilistic forecasts in terms of
predictive distributions $p_r$ for observations $y_r$ by
cross-validation, that is,~the data set is divided into subsets of
approximately equal size and predictions for one of the subsets are
obtained from estimates based on all the remaining subsets. Let
$y_{1},\ldots,y_{R}$ be data in a hold-out sample and $p_r$ the
predictive distributions with predicted parameter vectors $\hat
{\bolds{\vartheta}}_r=(\hat\vartheta_{r1},\ldots,\hat\vartheta_{rK})'$,
$r=1,\ldots,R$. Competing\vspace*{1pt} forecasts are then ranked by averaged scores
$S=\frac{1}{R}\sum_{r=1}^R S(p_r,y_r)$
such that higher scores deliver better probabilistic forecasts when
comparing different models.

\begin{figure}[b]

\includegraphics{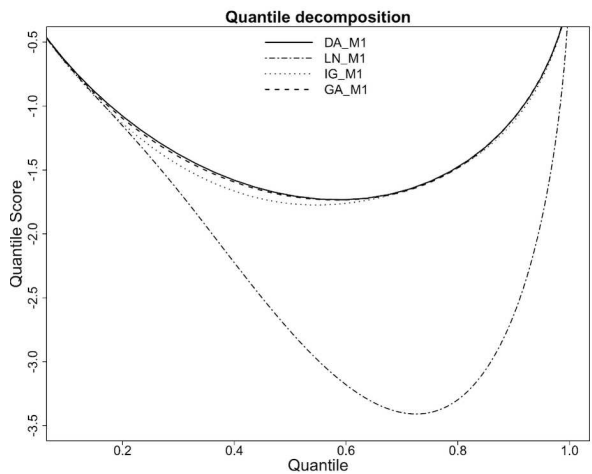}

\caption{Quantile decomposition of CRPS in the full
models DA\_M1, LN\_M1, IG\_M1, GA\_M1.}\label{threquaza}
\end{figure}

In our application, we conducted ten-fold cross-validation;
observations are assigned randomly to the different folds. The scores
discussed above are documented in Table~\ref{tabdic-scores-income}
where the values are averages of the ten folds (and scores within the
folds are themselves averages over the individual score contributions).
In line with the DIC and the residual plots, the scores of the DA\_M1
model are the highest and thus deliver the best forecast among the 12
models under consideration. Also similar to the DIC, models of type M2
(the simplest versions) show lower scores compared to the ones of type
M3 and they themselves are inferior compared to the most flexible
models of type M1.

In addition to the averages over the ten folds, the proper scoring
rules can also be used to assess the predictive distributions in more
detail. We illustrate this along a decomposition of the CRPS over
quantile levels (Figure~\ref{threquaza}) and a decomposition of the
scores over the cross-validation folds; compare the supplement
Section~A.1. The quantile level decomposition of
the CRPS again indicates a comparable performance of the Dagum
distribution and the gamma distribution as compared to the inverse
Gaussian distribution which performs somewhat worse and the log-normal
distribution which shows a considerably deteriorated behaviour. This
ordering holds true over the complete range of quantiles. The fact that
the log-normal distribution fails to provide a competing predictive
ability is most probably related to the strong impact of the extreme
observations. These are hard to capture by the log-normal distribution
in general. However, since extreme observations are typically also
influential observations, they seem to impact estimates in the
log-normal model to such an extent that even predictions for the
central part of the distribution are affected negatively.

\section{Regional disparities of the distribution of labour income in
Germany}\label{secincome}

As discussed in the \hyperref[intro]{Introduction}, our main focus is on investigating
differences in conditional income distributions between former East and
West Germany in the first decade of the new millennium. More
specifically, we focus on differences in the inequality of the
conditional income distribution as measured by the Gini coefficient
[\citet{Silber1999}] next to significant differences in the first two
moments of conditional income distributions. Based on our model choice,
we illustrate the estimation results along the Dagum model DA\_M1.

In their seminal paper, \citet{DiNardo1996},  stress the need
to look at differences between the whole conditional income
distributions rather than just the conditional mean income, or certain
indices. Using our proposed estimation procedure, this is feasible.
Figure~\ref{Figcids} displays an exemplary contrast of four
conditional income distributions in a ceteris paribus type analysis. %
The four distributions have all but two covariates fixed at their
average value. For age (42 years) and labour market experience (19
years) we use the arithmetic mean of the observations in our sample,
while we fixed the random effects at their prior expectation, that
is,~at zero.
Keeping these covariates fixed, we can observe the nature of the change
if the regional variable is changed from East to West. For both
educational levels, this figure furthermore indicates that there is a
noticeable difference not only in the mean value of the distributions
but also in other aspects, like variability, skewness, etc. Thus, a
simple analysis of means falls short of portraying a comprehensive
picture of the differences in income between East and West.

Note that for determining the densities displayed in Figure~\ref{Figcids} we consider the posterior mean of the densities obtained in
the different MCMC iterations instead of plugging in the posterior mean
parameters in the corresponding parametric densities. The availability
of such posterior mean estimates is another advantage of the Bayesian
inferential approach based on MCMC simulations.

There are various additional aspects of the distribution that can be
considered. In principle, it is possible to obtain any distributional
measure from the conditional distribution as long as it is defined for
the given distribution type and the corresponding parameter set.
Here, we consider the mean, the standard deviation and the Gini
coefficient of the estimated conditional income distributions. While
the mean provides important information on the location of the income
distribution, the standard deviation provides information on the scale
of the distribution and the Gini coefficient is the most frequently
used scalar measure on income inequality [\citet{Silber1999}]. We
therefore look at three important aspects of the conditional income
distributions and observe how they change over the covariate space.

\subsection{The spatial effect on conditional means and standard deviations}

\begin{figure}[b]

\includegraphics{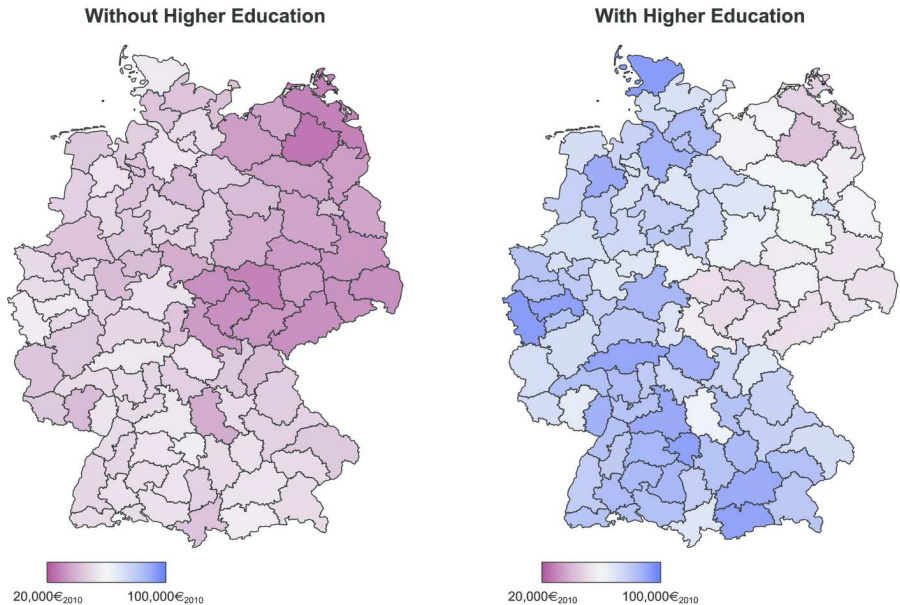}

\caption{SOEP data. Posterior means for the expected
incomes for 42-year-old males with 19 years of working experience.
Left: males without higher education. Right: males with higher
education.} \label{FigMeanLE}
\end{figure}

Assuming a Dagum distribution, the first two moments of the conditional
income distributions of $y_i$ can be found in \citet{Dagum2008},  respectively.
Figure~\ref{FigMeanLE} displays the posterior mean estimates for the
expected incomes for each of the 96 regions (\textit
{Raumordnungsregionen}) and education. As described above, the other
covariates are fixed at their mean.

Unsurprisingly, there is a clearly visible divide between East and
West Germany, as expected incomes are higher in the former Federal
Republic of Germany for both education levels and at the average of the
other covariates. Abstracting from the variations at the district
level, we get an expected income of 33{,}600{\euro} if the average man
lives in the East and has no higher education. With higher education
the income increases to 55{,}200{\euro}. The corresponding values if a
person with the same attributes lives in the West are 48{,}100{\euro} and
78,300{\euro}. The difference between East and West is thus
14,500{\euro}
(12{,}000{\euro}; 17{,}100{\euro}) and 23{,}100{\euro} (19{,}000{\euro};
27{,}400{\euro}) without and with higher education, respectively, where the numbers in
the brackets denote the corresponding 95\% credible intervals. In
addition to posterior means, we also looked at posterior medians.
Overall, differences were negligible, which is in line with the theory
suggesting asymptotic normality for the posterior distribution.

\begin{figure}[b]

\includegraphics{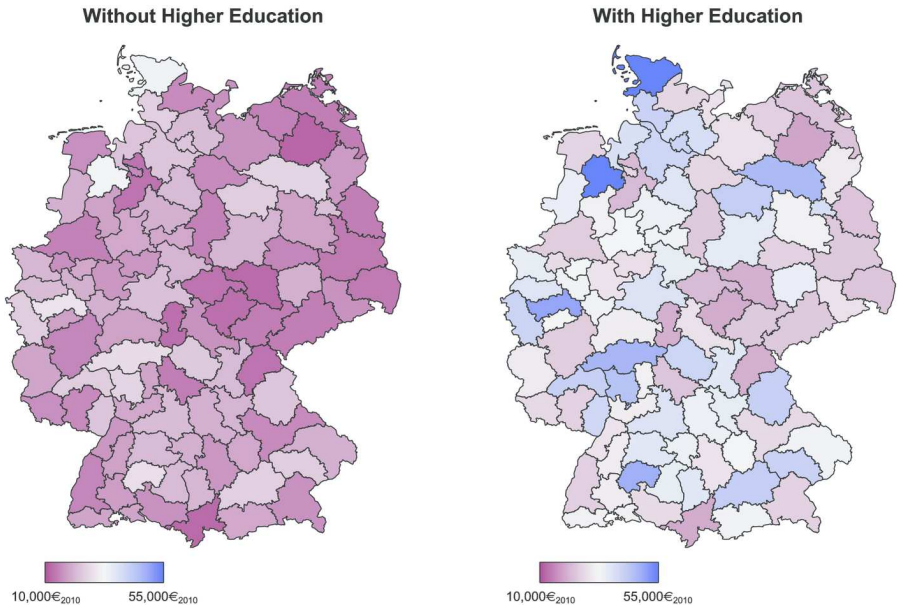}

\caption{SOEP data. Posterior means for standard
deviations for 42-year-old males with 19 years of working experience.
Left: males without higher education. Right: males with higher
education.} \label{FigsdLE}
\end{figure}
The posterior mean estimates for the standard deviations of the
conditional income distributions are shown in Figure~\ref{FigsdLE}. We
prefer presenting the square roots of the second moments, that is,~we
consider the standard deviations rather than the variances for
interpretability reasons.

For standard deviations, the division between East and West is not as
distinct as for the means. The main difference in the scale of the
conditional distributions is found between the education levels and not
along the different regions or former two parts of Germany.
Nonetheless, if we set the spatial random effect to zero again and only
consider the structural effect, the resultant conditional distribution
in the West has a standard deviation of 19{,}300{\euro}, while that of the
East has a standard deviation of 16{,}000{\euro}
for those without higher
education. For those with higher education the respective numbers are
32{,}000{\euro} and 26{,}600{\euro}. The difference between the standard
deviations is thus 3300{\euro} (1300{\euro}; 5200{\euro}) in the group of
lower educated males and 5400{\euro} (1700{\euro}; 9100{\euro}) for the one
with higher educated males.

Our results show that evaluated at the mean of other covariates, the
first and second moment are significantly different in East
and West Germany for both education levels, highlighting the diverse
nature of the change of conditional income distributions.

\subsection{The spatial effect on the conditional income inequality}
The Gini coefficient is an inequality measure based on the Lorenz curve
[\citet{Sarabia2008}], which can vary between the value 0 (everybody
has the same) and 1 (one person has everything). Note that the Gini
coefficient is scale invariant such that in standard mean regression on
log-incomes it would be postulated as constant across the covariate space.
In analogy to the conditional mean income and standard deviation, the
Gini coefficient of the conditional income distribution can easily be
obtained from the parameter estimates of the Dagum
distribution~[\citet{Dagum2008}, page 104].

\begin{figure}[b]

\includegraphics{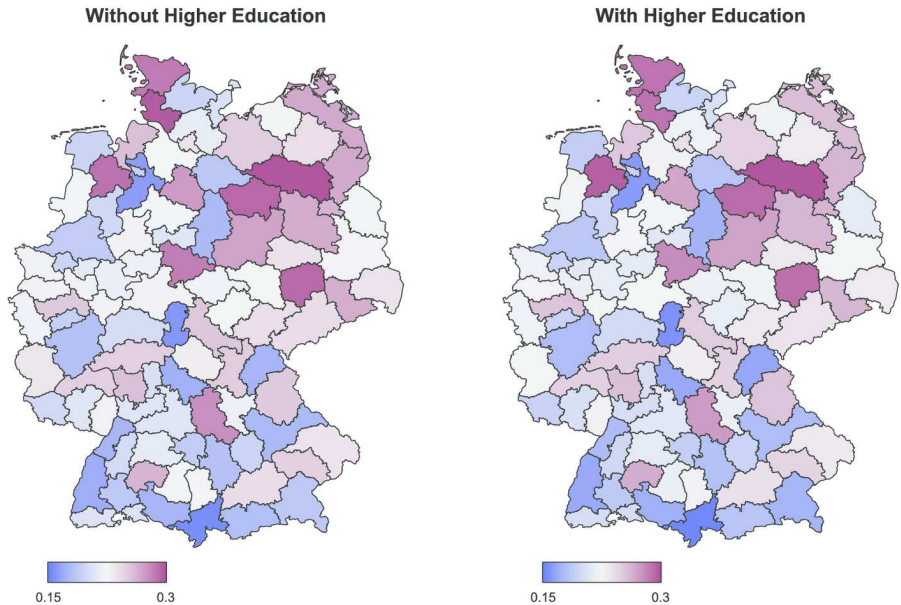}

\caption{SOEP data. Posterior means for the Gini
coefficients for 42-year-old males with 19 years of working experience.
Left: males without higher education. Right: males with higher
education.} \label{FigGini}
\end{figure}

Figure~\ref{FigGini} portrays the posterior mean estimates for the
Gini coefficients for each region. As we can see, the differences are
not as clear cut as for the conditional mean incomes. Nonetheless, the
pattern emerging indicates that income inequality among 42-year-old
males with 19 years of experience is higher in the East for both
education levels. Indeed, if we only consider the impact of the binary
East--West variable on the Gini coefficient, we obtain a difference of
the posterior means of 0.039 and 0.036 for those without higher
education and those with higher education, respectively. The
corresponding 95\% credible intervals are [$0.015, 0.067$] and
[$0.013, 0.063$], respectively. Thus, we have a significantly larger
income inequality for 42-year-old males with 19 years of experience, as
measured by the Gini coefficient, in the East than in the West. Putting
these differences into perspective, the standard deviations of the Gini
coefficients of the regions' conditional income distributions within
East and West are 0.030 and 0.031 for those without higher education,
and 0.032 and 0.031 for those with higher education. Thus, the
differences between East and West are not only significant, they also
surpass their variation within East and West.


\subsection{Further analysis of the conditional income distribution}

\subsubsection*{The effect of varying age and experience}
Next to spatial effects, the impact of the other covariates can also be
of interest. In the following, we focus on the effects of age and
experience, while the effect of year is treated in Section~A.1.1. For
results on additional covariate sets, see Section~A.1.2.

%
\begin{figure}[b]

\includegraphics{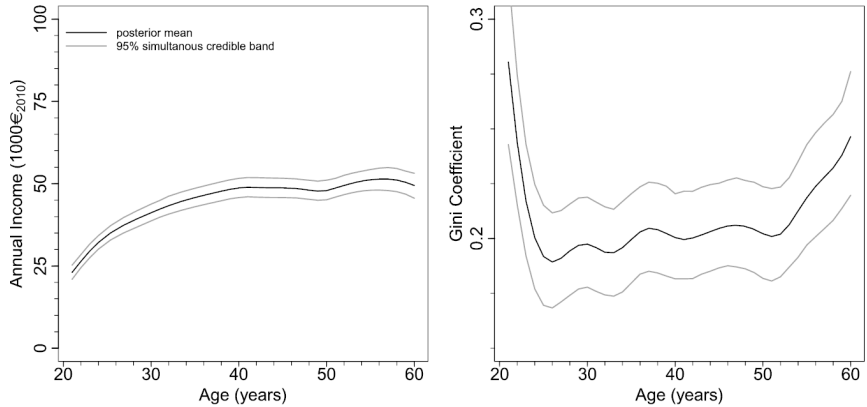}

\caption{SOEP data. Posterior means for expected income
(left) and Gini coefficients (right) for males who have been working
since the age of 21, without higher education and living in the West,
together with 95\% simultaneous credible bands.} \label{Figmeanage}
\end{figure}

In Figure~\ref{Figmeanage}, we display the expected conditional mean
income and the Gini coefficient with respect to age and experience. In
order to keep the dimension of the varying covariate to one,
we simply assume that from the age of 21 onwards people gain one year
of work experience as they grow older by one year. Here, we thus
portray the development of expected incomes and the Gini coefficient
for full-time working males who have been working since the age of 21.
With regard to the categorical variables region and education, we
consider only the West and lower education, respectively. The random
effects for the \textit{Raumordnungsregion} and year are considered at
zero, that is,~their prior expectation. The grey lines indicate the 95\%
simultaneous credible bands.
As expected, there is a general upward trend such that expected incomes
are rising with increasing age. In addition, we see the concave
structure that is generally also found by the literature.

For the Gini coefficient,  we observe a U-shaped development over age.
This indicates that the conditional income distribution is not simply
rescaled over the age range but rather that it changes its shape such
that the Gini coefficient rises. Our results are again in line with
economic theory. At the very beginning of the career, income
inequalities should be rather high, as large parts are still not yet
allocated in accordance to their capabilities and, consequently, are
employed and paid more or less arbitrarily. These mismatch-induced
inequalities quickly fade away. From then on we would expect rising
inequality, as following the classical theories on the shape of the
unconditional income distribution [\citet{Arnold2008}]; the
latter is
made up of incomes derived from a varying number of autoregressive
permutations. These permutations, which would generally occur over the
age range under consideration, would lead to a rising inequality in
incomes with rising age.

\subsubsection*{Other quantities derived from conditional income distributions}
Using distributional regression, it is easily possible to obtain
estimates for certain quantiles, like the median, which is an
alternative to the mean as a location measure. Furthermore, one can
calculate interquantile ranges as an alternative measure of inequality.
Naturally, such quantiles can be estimated in a more direct manner
using quantile regression, although additional efforts may be required
to avoid crossing quantile curves, in particular when considering a
dense set of quantile levels. Distributional regression automatically
avoids the problem of quantile crossing and makes model comparison
easier in such situations. 
We contrast distributional regression against quantile regression in
more detail for our case study in Section~A.1.3.

Next to measures of inequality like the Gini coefficient or the Theil
index, which are easily computable, it is also straightforward to
calculate measures of polarisation, which have recently received
considerable attention in the literature [for further references and
explanations, see, e.g., \citet{Wolfson1994,Duclos2004}]. Following
\citet{Gradin2000}, it would be possible to calculate the polarisation
between two groups as defined by sets of covariates.

It is also possible to assess density differences at different income
levels or probability mass differences for different income ranges. For
instance, one could consider the probability mass above a certain
income, for example,~48,000\euro, which according to John Keynes would
suffice to turn one's mind away from pecuniary worries [\citet
{Skidelsky2010}]. Consequently, it could be highlighted that not only
the conditional mean income for the average man without higher
education is lower in the East but also that the probability mass of
incomes below that threshold is much lower. Such an analysis may be of
particular interest for research questions on poverty and
vulnerability [\citet{Pudney1999}].

\subsection{Economic consequences}
Our findings show that keeping other variables fixed at their average
level, there are significant differences in income inequalities within
East and West Germany. \citet{Duclos2004} have noted the
importance of
within-group inequality for levels of alienation and identification
within society. The higher income inequality in the East would thereby
induce a weakened in-group identity. Lack of in-group identity in turn
is likely to cause feelings of isolation and mistrust [\citet
{Misztal2013}], and thus leads to a deterioration of well-being which
is beyond that captured by solely considering average incomes, or even
distribution-adjusted well-being measures [\citet{Klasen2008}].

While a profound analysis of the effect of different income
distributions to well-being must be left for further research, our
application shows that structured additive distributional regression
offers a methodology to the analysis of income \mbox{inequality} which goes
beyond the analysis at a highly aggregated level and thus allows to
start the assessment of this important issue at a microeconomic level.

\section{Conclusion}\label{secsummary}

Distributional regression and the closely related class of GAMLSS
provide a flexible, comprehensive toolbox for solving complex
regression problems with potentially nonstandard response types. They
are therefore useful to overcome the limitations of common mean
regression models and to enable a proper, realistic assessment of
regression relationships. In this paper, we provided a Bayesian
approach to distributional regression and described solutions for the
most important applied problems, including the selection of a suitable
predictor specification and the most appropriate response distribution.
Based on efficient MCMC simulation techniques, we developed a generic
framework for inference in Bayesian structured additive distributional
regression relying on distribution-specific iteratively weighted least
squares proposals as a core feature of the algorithms.

Concerning the specific application of distributional regression to
conditional income distributions, there are significant differences
between men with similar age, work experience and education levels
between East and West which go beyond the mean income. Taking the Gini
coefficient as an indicator for inequality, income inequality among
these men is larger in the East than it is in the West, further
deepening differences in well-being. While this study highlights the
scope of the new methodology to an application of income analysis and
beyond, much work remains to be done on the application of
distributional regression techniques.

Despite the practical solutions outlined in this paper, model choice
and variable selection remain relatively tedious and more automatic
procedures would be highly desirable.
Suitable approaches may be in the spirit of \citet{bellan08} in a
frequentist setting or based on spike and slab priors for Bayesian
inference as developed in \citet{schfah11} for mean regression.

It will also be of interest to extend the distributional regression
approach to the multivariate setting. For example, in the case of
multivariate Gaussian responses, covariate effects on the correlation
parameter may be very interesting in specific applications. Similarly,
multivariate extensions of beta regression lead to Dirichlet
distributed responses representing multiple percentages that sum up to
one; see \citet{KleKneKlaLan2013} for a first attempt in this direction.

In the context of economic applications, it should be noted that,
analogously to generalised linear models, the additive impact of
explanatory variables on the economic measure of interest, like the
Gini coefficient, is generally not attained. Consequently, the size,
and possibly also the direction of the estimated spatial effect, may
well be very different for different points in the covariate space.
While it is straightforward to calculate these differences with
corresponding credible intervals for any desired combination of other
covariates to give a more comprehensive assessment of differences in
inequality, further work needs to be done to facilitate the
interpretation of results.

In addition, in-depth-testing is required to find adequate parametric
forms for conditional income distributions, as the application of
structured additive distributional regression crucially rests on the
assumption that the parametric distribution fits the data. While for
the case of full-time working men the Dagum distribution indeed seems
to provide a decent fit, further work must be done to allow for an
analysis with a less restricted covariate space and thus a more
comprehensive analysis of income distributions in Germany and beyond.

Yet, this paper demonstrates that structured additive distributional
regression offers a statistical framework addressing the challenge to
assess entire conditional distributions [\citet{Fortin2011}, page 56]
by broadening the class of potential response distributions beyond
simple exponential families and thus offers additional scope for
applied statistical analyses on the problem of income inequality and beyond.

\section*{Acknowledgments}
We thank two referees, the Associate Editor
and Tilmann Gneiting for their careful review which was very helpful in
improving upon the initial submission.

\begin{supplement}[id=suppA]
\sname{Supplement A}
\stitle{Case studies}
\slink[doi]{10.1214/15-AOAS823SUPPA}
\sdatatype{.pdf}
\sfilename{AOAS823\_suppA.pdf}
\sdescription{Additional material on the application to regional
income inequality in Germany is provided in Section A.1. A second case
study on the proportion of farm outputs achieved by cereals is treated
in Section A.2.}
\end{supplement}

\begin{supplement}[id=suppB]
\sname{Supplement B}
\stitle{Methodology}
\slink[doi]{10.1214/15-AOAS823SUPPB}
\sdatatype{.pdf}
\sfilename{AOAS823\_suppB.pdf}
\sdescription{This supplement comprises details on Bayesian inference,
derivations of required quantities for the iteratively weighted least
squares proposals and simulation studies.}
\end{supplement}

%
%

\printaddresses
\end{document}